\documentclass[12pt]{article}

\usepackage{amsmath}

\usepackage[dvipdfmx]{graphicx}

\usepackage{subfig}

\usepackage{array}

\usepackage{amsfonts}

\usepackage{epsf}

\usepackage{epsfig}

\usepackage{amssymb}

\usepackage{bbm}

\usepackage{delarray}

\usepackage{caption}

\usepackage{amstext}

\usepackage{graphics}

\usepackage{epstopdf}

\usepackage[titletoc,toc,title]{appendix}

\usepackage{scrtime}

\usepackage[multiple]{footmisc}

\usepackage[square,numbers]{natbib}

\usepackage{color}

\makeatletter

\catcode`ð=\active
 \defð{\u{g}}
 \catcode`Ð=\active
 \defÐ{\u{G}}
 \catcode`Ý=\active
 \defÝ{\. I}
 \catcode`ö=\active
 \defö{\"{o}}
 \catcode`Ö=\active
 \defÖ{\"O}
 \catcode`ü=\active
 \defü{\"{u}}
 \catcode`Ü=\active
 \defÜ{\"{U}}
 \catcode`Þ=\active
 \defÞ{\c{S}}
 \catcode`þ=\active
 \defþ{\c{s}}
 \catcode`ý=\active
 \defý{{\i}}
 \catcode`ç=\active
 \defç{\d{c}}
 \catcode`Ç=\active
 \defÇ{\d{C}}

\renewcommand*\@fnsymbol[1]{\the#1}

\makeatother

\numberwithin{equation}{section}
\begin{document}

\title{\textbf{Thermodynamic properties of a charged particle in non-uniform magnetic field}}

\author{
{\small H. R. Rastegar Sedehi\footnote{h.rastegar83@yahoo.com}\,\,}\\
{\small \emph{Department of Physics, Jahrom University}}, \\
{\small \emph{74137-66171 Jahrom, Iran}}\\
{\small Altu\u{g} Arda\,\,}\\
{\small \emph{Division of Physics Education, Hacettepe University}}, \\
{\small \emph{06800, Ankara, Turkey}}\\
{\small Ramazan Sever\,\,}\\
{\small \emph{Department of Physics, Middle East Technical University}}, \\
{\small \emph{06800, Ankara, Turkey}}\\ \\}
\date{}

\maketitle

\begin{abstract}
We solve the Schrödinger equation for a charged particle in the non-uniform magnetic field by using the Nikiforov-Uvarov method. We find the energy spectrum and the wave function, and present an explicit relation for the partition function. We give analytical expressions for the thermodynamic properties such as mean energy and magnetic susceptibility, and analyze the entropy, free energy and specific heat of this system numerically. It is concluded that the specific heat and magnetic susceptibility increase with external magnetic field strength and different values of the non-uniformity parameter, $\alpha$, in the low temperature region, while the mentioned quantities are decreased in high temperature regions due to increasing the occupied levels at these regions. The non-uniformity parameter has the same effect with a constant value of the magnetic field on the behavior of thermodynamic properties. On the other hand, the results show that transition from positive to negative magnetic susceptibility depends on the values of non-uniformity parameter in the constant external magnetic field.

PACS: 03.65.-w, 03.65.Pm, 11.10.Wx
  %\narrowtext

Keywords: Exact Solution, Non-uniform Magnetic Field, Statistical Quantity, Specific heat,  Magnetic Susceptibility, Nikiforov-Uvarov Method
\end{abstract}
\newpage

\section{Introduction}
The Schrödinger equation is a partial differential equation that describes the dynamics of quantum mechanical systems via the wave function. It's solutions have important applications in condensed matter, atomic, nuclear, particle and high energy physics [1-4]. One of the methods used in the solution of the Schrödinger equation is the Nikiforov-Uvarov (NU) method [5].
The analytical and/or numerical solutions of the Schrödinger equation in a magnetic field have been investigated by many scientists in recent years [6-9].  Khordad et al. [10] have solved analytically the Schrödinger equation for the Morse ring-shaped potential by using the Pekeris approximation. They have used NU procedure to determine the thermodynamic properties of the $TiC$ diatomic molecule. The thermodynamic properties of real diatomic molecule or triatomic molecule systems have been predicted successfully in these Refs. [11-15]. Hassanabadi et al. [16] have solved the Schrödinger equation for an energy-dependent potential. They have calculated the eigenfunctions and eigenvalues for an energy-dependent Hamiltonian that linearly depends on energy in the general $D$-dimensional space with the help of NU method.

A number of authors have studied the effect of non-uniform magnetic fields on the properties of various systems [17-23]. Contreras-Astorga et al. [24] have investigated an electron under an external non-uniform magnetic field, where he shape-invariant properties of the reduced radial matrix Hamiltonian have been studied by using the discrete spectrum and eigenfunctions.

Among the physical properties, there are some studies on thermodynamic properties of different systems [25-34]. Tan et al. [35] have studied the magnetization and persistent currents in mesoscopic rings and dots by use of an exactly soluble model to calculate the structures of large number of electrons over a wide range of magnetic field strengths. Khordad et al. [36] have investigated the effects of the Rashba spin-orbit interaction and applied magnetic field on thermodynamic properties of a quasi-one-dimensional quantum wire at low temperatures. There are also some studies investigating thermal, and magnetic effects on other type of structures [37-43].

We will remark the most important thermodynamics properties of the charged particle in a magnetic field. We have solved the Schrödinger equation for this system in the non-uniform magnetic field by using the NU method and have found the energy spectrum and the wave function. We have presented an explicit relation for partition function by using Poisson summation formula. We give analytical expressions for the thermodynamics properties such as mean energy and magnetic susceptibility.

This work is organized as follows: In Sec. 2, we give a brief review of the NU method, which includes the required expressions for the present problem. In Sec. 3, we solve the Schrödinger equation for a charged particle in a non-uniform magnetic field to calculate the partition function, and then display the thermodynamics quantities such as entropy, internal energy, specific heat, magnetization and magnetic susceptibility. In Sec. 4, we summarize our discussions on the numerical results. We give conclusions in last section.

\section{Nikiforov-Uvarov Method}

We summarize here the new version of the Nikiforov-Uvarov method presented in Ref. [7]. Let us write the generalized form of the Schr\"{o}dinger-like equation written for any
potential as

\begin{eqnarray}
\left[\frac{d^{2}}{ds^{2}}+\frac{\alpha_{1}-\alpha_{2}s}{s(1-\alpha_{3}s)}
\frac{d}{ds}+\frac{-\xi_{1}s^{2}+\xi_{2}s-\xi_{3}}{[s(1-\alpha_{3}s)]^{2}}\right]\Psi(s)=0.
\end{eqnarray}
Comparing it with the basic differential equation used in the standard NU-method gives
\begin{eqnarray}
\tilde{\tau}(s)=\alpha_{1}-\alpha_{2}s\,\,\,;\,\,\sigma(s)=s(1-\alpha_{3}s)\,\,\,;\,\,
\tilde{\sigma}(s)=-\xi_{1}s^{2}+\xi_{2}s-\xi_{3}\,.
\end{eqnarray}
The function $\pi(s)$ becomes
\begin{eqnarray}
\pi(s)=\alpha_{4}+\alpha_{5}s\pm\sqrt{(\alpha_{6}-k\alpha_{3})s^{2}+(\alpha_{7}+k)s+\alpha_{8}}\,,
\end{eqnarray}
where the parameter set are
\begin{eqnarray}
\begin{array}{ll}
\alpha_{4}=\frac{1}{2}\,(1-\alpha_{1})\,, & \alpha_{5}=\frac{1}{2}\,(\alpha_{2}-2\alpha_{3})\,, \\
\alpha_{6}=\alpha_{5}^{2}+\xi_{1}\,, &
\alpha_{7}=2\alpha_{4}\alpha_{5}-\xi_{2}\,, \\
\alpha_{8}=\alpha_{4}^{2}+\xi_{3}\,. & \\
\end{array}
\end{eqnarray}
The function under the square root in Eq. (2.3) is to
be the square of a polynomial [5]. This condition gives the roots
of the parameter $k$ written as
\begin{eqnarray}
k_{1,2}=-(\alpha_{7}+2\alpha_{3}\alpha_{8})\pm2\sqrt{\alpha_{8}\alpha_{9}}\,,
\end{eqnarray}
where the $k$-values can be real or imaginary, and
$\alpha_{9}=\alpha_{3}\alpha_{7}+\alpha_{3}^{2}\alpha_{8}+\alpha_{6}$\,.
It is clear that $k_1$ and $k_2$ in Eq. (2.5) lead to different $\pi(s)$-functions in Eq. (2.3). For
\begin{eqnarray}
k=-(\alpha_{7}+2\alpha_{3}\alpha_{8})-2\sqrt{\alpha_{8}\alpha_{9}}\,,
\end{eqnarray}
$\pi(s)$ becomes
\begin{eqnarray}
\pi(s)=\alpha_{4}+\alpha_{5}s-\left[(\sqrt{\alpha_{9}}+\alpha_{3}\sqrt{\alpha_{8}}\,)s-\sqrt{\alpha_{8}}\,\right]\,,
\end{eqnarray}
and also
\begin{eqnarray}
\tau(s)=\alpha_{1}+2\alpha_{4}-(\alpha_{2}-2\alpha_{5})s-2\left[(\sqrt{\alpha_{9}}
+\alpha_{3}\sqrt{\alpha_{8}}\,)s-\sqrt{\alpha_{8}}\,\right].
\end{eqnarray}
It should be imposed here the following expression for satisfying the condition that
the derivative of the function $\tau(s)$ should be negative in the
method
\begin{eqnarray}
\tau^{\prime}(s)&=&-(\alpha_{2}-2\alpha_{5})-2(\sqrt{\alpha_{9}}+\alpha_{3}\sqrt{\alpha_{8}}\,)
\nonumber \\
&=&-2\alpha_{3}-2(\sqrt{\alpha_{9}}+\alpha_{3}\sqrt{\alpha_{8}}\,)\quad<0.
\end{eqnarray}
In this approach, the following equation is the eigenvalue equation [7]
\begin{eqnarray}
\alpha_{2}n-(2n+1)\alpha_{5}&+&(2n+1)(\sqrt{\alpha_{9}}+\alpha_{3}\sqrt{\alpha_{8}}\,)+n(n-1)\alpha_{3}\nonumber\\
     &+&\alpha_{7}+2\alpha_{3}\alpha_{8}+2\sqrt{\alpha_{8}\alpha_{9}}=0.
\end{eqnarray}
The weight function $\rho(s)$ in NU-method is written here as
\begin{eqnarray}
\rho(s)=s^{\alpha_{10}-1}(1-\alpha_{3}s)^{\frac{\alpha_{11}}{\alpha_{3}}-\alpha_{10}-1}\,,
\end{eqnarray}
so we have
\begin{eqnarray}
y_{n}(s)=P_{n}^{(\alpha_{10}-1,\frac{\alpha_{11}}{\alpha_{3}}-\alpha_{10}-1)}(1-2\alpha_{3}s)\,,
\end{eqnarray}
where
\begin{eqnarray}
\alpha_{10}=\alpha_{1}+2\alpha_{4}+2\sqrt{\alpha_{8}}\,\,;\,
\alpha_{11}=\alpha_{2}-2\alpha_{5}+2(\sqrt{\alpha_{9}}+\alpha_{3}\sqrt{\alpha_{8}})\,.
\end{eqnarray}
and $P_{n}^{(\alpha,\beta)}(1-2\alpha_{3}s)$ are the Jacobi
polynomials. The other part of the general solution is given
\begin{eqnarray}
\phi(s)=s^{\alpha_{12}}(1-\alpha_{3}s)^{-\alpha_{12}-\frac{\alpha_{13}}{\alpha_{3}}}\,,
\end{eqnarray}
then the general solution $\Psi(s)=\phi(s)y(s)$ becomes
\begin{eqnarray}
\Psi(s)=s^{\alpha_{12}}(1-\alpha_{3}s)^{-\alpha_{12}-\frac{\alpha_{13}}{\alpha_{3}}}
P_{n}^{(\alpha_{10}-1,\frac{\alpha_{11}}{\alpha_{3}}-\alpha_{10}-1)}(1-2\alpha_{3}s)\,,
\end{eqnarray}
where
\begin{eqnarray}
\alpha_{12}=\alpha_{4}+\sqrt{\alpha_{8}}\,\,;\,\alpha_{13}=\alpha_{5}-(\sqrt{\alpha_{9}}+\alpha_{3}\sqrt{\alpha_{8}}\,)\,.
\end{eqnarray}
If the parameter $\alpha_3$ is zero in the problem then $\Psi(s)$ in Eq. (2.15) turns into [5, 7]
\begin{eqnarray}
\Psi(s)=s^{\alpha_{12}}e^{\alpha_{13}s}L_{n}^{\alpha_{10}-1}(\alpha_{11}s)\,,
\end{eqnarray}
when
\begin{eqnarray}
\lim_{\alpha_3 \rightarrow 0}P_{n}^{(\alpha_{10}-1,\frac{\alpha_{11}}{\alpha_{3}}-\alpha_{10}-1)}(1-2\alpha_{3}s)=L_{n}^{\alpha_{10}-1}(\alpha_{11}s)\,,
\end{eqnarray}
and
\begin{eqnarray}
\lim_{\alpha_3 \rightarrow 0}(1-\alpha_{3}s)^{-\alpha_{12}-\frac{\alpha_{13}}{\alpha_{3}}}=e^{\alpha_{13}s}\,.
\end{eqnarray}
For the other value of the parameter $k$ giving
\begin{eqnarray}
k=-(\alpha_{7}+2\alpha_{3}\alpha_{8})+2\sqrt{\alpha_{8}\alpha_{9}}\,,
\end{eqnarray}
the corresponding solution for the wave function and the eigenvalue equation are given by [7]
\begin{eqnarray}
\Psi(s)=s^{\alpha^{*}_{12}}(1-\alpha_{3}s)^{-\alpha^{*}_{12}-\frac{\alpha^{*}_{13}}{\alpha_{3}}}
P_{n}^{(\alpha^{*}_{10}-1,\frac{\alpha^{*}_{11}}{\alpha_{3}}-\alpha_{10}-1)}(1-2\alpha_{3}s)\,,
\end{eqnarray}
and
\begin{eqnarray}
\alpha_{2}n-2\alpha_{5}n&+&(2n+1)(\sqrt{\alpha_{9}}-\alpha_{3}\sqrt{\alpha_{8}}\,)+n(n-1)\alpha_{3}\nonumber\\
     &+&\alpha_{7}+2\alpha_{3}\alpha_{8}-2\sqrt{\alpha_{8}\alpha_{9}}+\alpha_5=0\,,
\end{eqnarray}
where
\begin{eqnarray}
\alpha^{*}_{10}&=&\alpha_1+2\alpha_4-2\sqrt{\alpha_8\,}\,,\nonumber\\
\alpha^{*}_{11}&=&\alpha_2-2\alpha_5+2(\sqrt{\alpha_9\,}-\alpha_{3}\sqrt{\alpha_{8}\,})\,,\nonumber\\
\alpha^{*}_{12}&=&\alpha_4-\sqrt{\alpha_8\,}\,,\nonumber\\
\alpha^{*}_{13}&=&\alpha_5-(\sqrt{\alpha_9\,}-\alpha_{3}\sqrt{\alpha_{8}\,})\,.
\end{eqnarray}

\section{Model}
The Schrödinger equation for a charged particle with mass $m$, and electric charge $e$ in a magnetic field is written as
\begin{eqnarray}
\left[\frac{1}{2m}\left(-i\hbar\vec{\nabla\,}-\frac{e}{c}\vec{A\,}\right)^2+U(\vec{r\,})\right]\psi(\vec{r\,})=E\psi(\vec{r\,})\,,
\end{eqnarray}
where $U(\vec{r\,})$ is the potential energy, $c$ is speed of light, and $E$ total energy of the particle [44].

Here we set the vector potential $\vec{A\,}$ in Eq. (3.1) as
\begin{eqnarray}
A_y(x)=\frac{B_0}{\alpha}\,(1-e^{-\alpha x})\,,
\end{eqnarray}
which gives a non-uniform magnetic field in the $\hat{z}$-direction as $\vec{B\,}=B_0e^{-\alpha x}\hat{z}$, where $B_0$ is a constant, and the parameter $\alpha$ is the non-uniformity parameter [45]. This form of magnetic field can have a role in different branches of physics, like in chiral magnetic effect in particle physics [46], searching the quantum structure of graphene [47], and within the supersymmetric quantum mechanics where supersymmetry is broken [48].
In order to have also an eigenfunction of $y$- and $z$-component of the linear momentum operator, we write the wave function as
\begin{eqnarray}
\psi(\vec{r\,})=e^{\{i(k_y y+k_z z)\}}\phi(x)\,,
\end{eqnarray}
Inserting it into Eq. (3.1), and using the vector potential given in Eq. (3.2) gives
\begin{eqnarray}
\frac{1}{2m}\left\{-\hbar^2\,\frac{d^2}{dx^2}+\hbar^2k^2_z+\left[\hbar k_y-\frac{eB_0}{\alpha c}(1-e^{-\alpha x})\right]^2\right\}\phi(x)=E\phi(x)\,,
\end{eqnarray}
where we set $U(\vec{r\,})=0$. The term $\frac{\hbar^2k^2_z}{2m}$ is the kinetic energy in the z-direction which means that the particle moves freely and the motion in the $z$-direction is unaffected by the magnetic field. The total energy is the kinetic energy in the $z$-direction plus the energy of the particle in the $xy$-plane, $E=\frac{\hbar^2k^2_z}{2m}+E'$, where $E'$ is the energy of the particle in the plane. So, we write the Schrödinger equation for the motion in the plane as
\begin{eqnarray}
\left\{\frac{d^2}{dx^2}-\left[k_y-\frac{eB_0}{\alpha\hbar c}(1-e^{-\alpha x})\right]^2+2mE'\right\}\phi(x)=0\,,
\end{eqnarray}

Defining a new variable $s=e^{-\alpha x}$ turns Eq. (3.5) into
\begin{eqnarray}
\frac{d^2\phi(s)}{ds^2}+\frac{1}{s}\frac{d\phi(s)}{ds}+\frac{1}{s^2}\left(-\xi_1-\xi_2s-\xi_3s^2\right)\phi(s)=0\,,
\end{eqnarray}
where
\begin{eqnarray}
\xi_1&=&\frac{1}{\alpha^2}\left[-k^2_y+\frac{2mE'}{\hbar^2}+\frac{eB_0}{\alpha\hbar c}\left(2k_y-\frac{eB_0}{\alpha\hbar c}\right)\right]\,,\nonumber\\
\xi_2&=&\frac{2eB_0}{\alpha^3\hbar c}\left(k_y-\frac{eB_0}{\alpha\hbar c}\right)\,,\nonumber\\
\xi_3&=&\left(\frac{eB_0}{\alpha^2\hbar c}\right)^2\,.
\end{eqnarray}

Comparing Eq. (3.6) with Eq. (2.1), and with the help of Eqs. (2.17) and (2.22), we find the eigenfunctions
\begin{eqnarray}
\phi(s) \sim s^{\sqrt{\xi_1\,}}e^{-s\sqrt{\xi_3\,}}L_{n}^{2\sqrt{\xi_1\,}}(2\sqrt{\xi_3\,}s)\,,
\end{eqnarray}
where $L_n^{\eta}(x)$ are the associated Laguerre polynomials [49], and energy eigenvalues
\begin{eqnarray}
E'_n=\frac{\alpha^2\hbar^2}{2m}\left\{\left(n+\frac{1}{2}\right)\left[n+\frac{1}{2}+\frac{2}{\alpha}\left(k_y-\frac{eB_0}{\alpha\hbar c}\right)\right]\right\}\,.
\end{eqnarray}

In order to get the thermodynamical functions, it is required to calculate the partition function given as [50-54]
\begin{eqnarray}
Q=\sum_{n=0}^{N_{max}}e^{-\beta E'_{n}}\,,
\end{eqnarray}
where $\beta=1/k_BT$ with the Boltzmann constant $k_B$, and $E'_{n}$ are energy eigenvalues. We indicate here that the quantum thermodynamics has also the aim of writing the quantum formulation of thermodynamics far from equilibrium for nanoscale quantum systems, where particle number is much less than the order of $10^{23}$, even for single-particle systems coupled with a reservoir. The starting point of the subject is writing the Hamiltonian as $H=H_S+H_R+H_I$ where $H_S$ is the Hamiltonian of the system, and $H_R$ of the reservoir, $H_I$ denotes the interaction between them [55].

Substituting Eq. (3.9) into Eq. (3.10), we reach to
\begin{eqnarray}
Q=\sum_{n=0}^{N_{max}}e^{-\frac{\beta\alpha^2\hbar^2}{2m}(n+\frac{1}{2})\left[n+\frac{1}{2}+\frac{2}{\alpha}\left(k_y-\frac{eB_0}{\alpha\hbar c}\right)\right]}\,,
\end{eqnarray}
We use the following expression for a finite summation with the upper bound to evaluate the partition function in Eq. (3.11), the Poisson summation formula which is used under the lowest order approximation reads as [50-54]:
\begin{eqnarray}
\sum_{n=0}^{N}\,f(n)=\frac{1}{2}\left[f(0)+f(N+1)\right]+\int_{0}^{N+1}f(x)dx\,,
\end{eqnarray}
which gives us
\begin{eqnarray}
Q=\frac{1}{2}\left[e^{-\frac{1}{4}\,\beta C_0(1+2C_1)}+e^{-\frac{1}{2}\,\beta C_0(2N+3)(2N+3+2C_1)}\right]\nonumber\\
+\int_{0}^{N+1}e^{-\frac{1}{2}\,\beta C_0(2x+1)(2x+1+2C_1)}dx\,,
\end{eqnarray}
where
\begin{eqnarray}
C_0=\frac{\alpha^2\hbar^2}{2m}\,;\,\,\,C_1=\frac{2}{\alpha}\left(k_y-\frac{eB_0}{\alpha\hbar c}\right)\,.
\end{eqnarray}
The partition function in Eq. (3.13) can be written in terms of the imaginary Error function as
\begin{eqnarray}
Q&=&\frac{1}{2}\left[e^{-\frac{1}{4}\beta C_0(1+2C_1)}+e^{-\frac{1}{2}\beta C_0(2N+3)(2N+3+2C_1)}\right]\nonumber\\
&+&\frac{1}{2}\sqrt{\frac{\pi}{2\beta C_0}\,}e^{\beta C_0C^2_1/2}\left\{Erfi\left[{\sqrt{\frac{\beta C_0}{2}\,}}(2N+3+C_1)\right]-Erfi\left[{\sqrt{\frac{\beta C_0}{2}\,}}(1+C_1)\right]\right\}\,,\nonumber\\
\end{eqnarray}
where $Erfi(x)$ is defined as [49]
\begin{eqnarray}
Erfi(x)=-iErf(ix)=\sqrt{\frac{4}{\pi}\,}\int_{0}^{x}e^{t^2}dt\,.
\end{eqnarray}

With the help of the partition function in Eq. (3.15), one can now derive the thermodynamic functions as
\begin{description}
  \item[(i)] Mean energy, $U=-\left[\frac{\partial}{\partial \beta}\,\ln Q\right]=\frac{\Lambda_1}{\Lambda_2}$

  here
  \begin{eqnarray}
  \Lambda_1&=&\frac{1}{4}\,C_0(2N+3)(C_1+D_1)e^{-\frac{1}{2}\,\beta C_0(2N+3)(C_1+D_1)}+\frac{1}{8}\,C_0(C_1+D_2)e^{-\frac{1}{4}\,\beta C_0(C_1+D_2)}\nonumber\\&-&\frac{1}{4}\frac{1}{\sqrt{2\beta C_0\,}}\left(\sqrt{\pi\,}C_0C^2_1e^{\frac{1}{2}\beta C_0C^2_1}(\Upsilon_1-\Upsilon_2)\right)-\frac{1}{\sqrt{2\beta C_0\,}}\left[e^{\frac{1}{2}\beta C_0C^2_1}( \right. \nonumber \\&\times& \left. e^{\frac{1}{2}\beta C_0D^2_1}(\frac{1}{2}\frac{C_0D_1}{\sqrt{\beta C_0\,}}-\frac{1}{4}\frac{\beta C^2_0D_1}{(\beta C_0)^{3/2}})\right. \nonumber \\ &-& \left. e^{-\frac{1}{2}\beta C_0D^2_2}(\frac{1}{2}\frac{C_0D_2}{\sqrt{\beta C_0\,}}-\frac{1}{4}\frac{\beta C^2_0D_2}{(\beta C_0)^{3/2}}))+\frac{1}{4}\,\sqrt{\frac{\pi}{2(\beta C_0)^3}\,}e^{\frac{1}{2}\beta C_0C^2_1}C_0(\Upsilon_1-\Upsilon_2)\right]\,,\nonumber\\
  \end{eqnarray}
  \begin{eqnarray}
  \Lambda_2&=&\frac{1}{2}\,e^{-\frac{1}{2}\beta C_0(2N+3)(C_1+D_1)}+\frac{1}{2}\,e^{-\frac{1}{4}\beta C_0\,(C_1+D_2)}+\frac{1}{2\sqrt{2\beta C_0\,}}\left[\sqrt{\pi\,}e^{-\frac{1}{4}\beta C_0C^2_1}\left(\Upsilon_1-\Upsilon_2\right)\right]\,,\nonumber\\
  \end{eqnarray}
  where
  \begin{eqnarray}
  &&\Upsilon_1=Erfi\left[\sqrt{\frac{\beta C_0}{2}\,}D_1\right]\,\,;\,\Upsilon_2=Erfi\left[\sqrt{\frac{\beta C_0}{2}\,}D_2\right]\,,\nonumber\\
  &&D_1=2N+C_1+3\,\,;D_2=C_1+1\,.\nonumber
  \end{eqnarray}
  \item[(ii)] Specific heat, $C=-\frac{\partial U}{\partial \beta}$
  \item[(iii)] Free energy, $F=-k_BT\ln Q$
  \item[(iv)] Entropy, $S=k_BT\ln Q-k_B\beta \frac{\partial \ln Q}{\partial \beta}$
  \item[(v)] Magnetic susceptibility, $\chi=-\frac{\partial^2 F}{\partial \beta^2}=-\frac{\Lambda_3}{\Lambda_4}+(\frac{\Lambda_5}{\Lambda_6})^2$

  with
  \begin{eqnarray}
  \Lambda_3&=&\frac{D_3(2N+3)^4}{2D^2_4\alpha^2}e^{-\frac{D_3}{4}(2N+3)(C_1+D_1)}+\frac{D^2_3}{8D^2_4\alpha^2}e^{-\frac{D_3}{8}(C_1+D_2)}\nonumber\\&+&\frac{\sqrt{\pi D_3\,}}{D^2_4\alpha^2}\,e^{D_5}(\Upsilon_3-\Upsilon_4)-\frac{\sqrt{\pi\,}C^2_1D^{3/2}_3}{2D^2_4\alpha^6}\,e^{D_5}(\Upsilon_3-\Upsilon_4)\,,
  \end{eqnarray}
  \begin{eqnarray}
  \Lambda_4=\frac{\beta}{2}\,\left[e^{-\frac{D_3}{8}(C_1+D_2)}+e^{-\frac{D_3}{2}(2N+3)(N+C_1)}+\frac{\sqrt{\pi\,}}{D_3}\,e^{D_5}(\Upsilon_3-\Upsilon_4)\right]\,,
  \end{eqnarray}
  \begin{eqnarray}
  \Lambda_5&=&\frac{1}{2D_4\alpha}\left[\frac{D_3}{2}\,e^{-\frac{D_3}{8}(C_1+D_2)}+D_3(2N+3)e^{-\frac{D_3}{4}(2N+3)(C_1+D_1)}\right. \nonumber \\ &-&\left. \sqrt{\pi D_3\,}C_1e^{D_5}(\Upsilon_3-\Upsilon_4)+2e^{D_5}\left(e^{\frac{1}{4}D_1D^2_2}-e^{\frac{1}{4}D_3D^2_1}\right)\right]\,,
  \end{eqnarray}
  \begin{eqnarray}
  \Lambda_6=\frac{\beta}{2}\,\left[e^{-\frac{D_3}{4}(2N+3)(C_1+D_1)}+e^{-\frac{D_3}{8}(C_1+D_2)}+\frac{\sqrt{\pi\,}}{D_3}\,e^{D_5}(\Upsilon_3-\Upsilon_4)\right]\,,
  \end{eqnarray}
  where
\begin{eqnarray}
&&\Upsilon_3=Erfi\left[\frac{1}{2}\,D_1\sqrt{D_3\,}\right]\,;\, \Upsilon_4=Erfi\left[\frac{1}{2}D_2\sqrt{D_3\,}\right]\,,\nonumber\\
&&D_3=\frac{\hbar^2\alpha^2\beta}{m}\,;\,D_4=\frac{\hbar c\alpha}{e}\,;\,D_5=\frac{\hbar^2\beta}{m}\left(k_y-\frac{B_0}{D_2}\right)^2\,.\nonumber
\end{eqnarray}
\end{description}

\section{Numerical Results}

In this part, we plot the numerical calculations for a charged particle in the non-uniform magnetic field. We investigate the effect of temperature and external magnetic field on thermodynamics properties of these system such as mean energy, entropy, specific heat, free energy, and magnetic susceptibility.

The mean energy as a function of temperature for different magnetic fields are shown in Fig. 1. It is seen that the mean energy increases slowly with temperature for different magnetic fields. When the magnetic fields increase, the energy levels splitting increases and therefore the mean energy depends on the distribution of energy levels and occupation probability of energy levels will be enhanced.

In Fig. 2 we study the changing of specific heat for a charged particle in the non-uniform magnetic field as a function of temperature for different magnetic fields with constant $\alpha$ parameter. In every curve, we see an anomalous peak the so-called Schottky anomaly that appears for different magnetic fields with respect to the temperature. Also, we see that the specific heat increases until a maximum value and then it reduces with enhancing the temperature. The summit of these curves goes to a more minor point with decreasing the magnetic field. At large temperatures the specific heat goes to constant value because it depends on the energy level distribution and the temperature dependence of the occupation probability of the states.

In Fig. 3 we plot the behaviour of the magnetic susceptibility as a function of the temperature for three different magnetic fields. The result shown that the magnetic susceptibility as a function of the temperature has positive values for different magnetic fields. Also, we have not any transition from a positive value to a negative one for the magnetic susceptibility. Remarkable in every curve is an anomalous peak which appears over a small range of temperatures and then all these curves go to a constant value increasing temperature.

Fig. 4 displays the free energy versus temperature for three different values of $\alpha$ parameter. It is obvious from the figure that the free energy increases until it reaches a maximum and then decreases with increasing temperature and goes to negative values. At a constant temperature the absolute value of free energy has higher value at smaller $\alpha$ parameter.

Fig. 5 shows the entropy as a function of the temperature for different $\alpha$ parameter. As we expect, the entropy increases with raising the temperature at a fixed $\alpha$ parameter. At fixed values of temperature, the entropy decreases with increasing $\alpha$ parameter due to the increase of the non-uniform magnetic field. Thereby the reduction of the system is disorder. With changing the value of the $\alpha$ parameter we can check the effects of different magnetic field on the system. It means that it does not need to change the value of the magnetic field for investigating the effect of different magnetic field. The different $\alpha$ parameters have the same effect with a constant value of the magnetic field on the behaviour of thermodynamic properties. It is necessary to notice that the different $\alpha$ parameters which leads to different values of magnetic field.

The variations of specific are plotted in Fig. 6 as a function of the temperature for different values of $\alpha$ parameter. The maximum value of the specific heat depends on the $\alpha$ parameter and decreases with raising this parameter. The Schottky anomaly as an interesting effect can be explained in terms of the change in the entropy of the system. At zero temperature the entropy is equal to zero because only the lowest level is occupied and there is a very little probability of transition to a higher energy level. With increasing the temperature, the entropy also increases and therefore the probability of the transition goes up. When all levels are occupied at high temperatures, there is a little change in the entropy for small changes in temperature and thus a lower heat capacity. Actually, by increasing the temperature, the probability of transition increases up to critical $T$ [called Schottky anomaly] and after that, due to the fully filled states, specific heat decreases with $T$. This critical temperature changes with entropy and interactions.

Fig. 7 displays the magnetic susceptibility obtained as a function of the temperature for three different values of $\alpha$ parameter. This quantity is calculated using the partition function and free energy. There is an interesting behaviour of magnetic susceptibility, it has only positive values for all used temperatures, and shows no transition temperature from positive values to negative values for choosen $\alpha$ parameter. In magnetic susceptibility the maximum point of the curves for different   changes with temperature. For our system, magnetic susceptibility shows an antiferromagnetic phase because the magnetic susceptibility starts from low values and increases up to a critical temperature, and then goes down. Also, the system does not demonstrate magnetic phase transition yet. It is necessary to mention that both specific heat and magnetic susceptibility quantities approach zero at high temperatures. The reason can be understood from the competition between the quantum effects and thermal effects. At high temperatures, thermal effects are dominant and there is no response for the quantum phenomena in them.

\section{Conclusions}
We have studied theoretically the thermodynamic properties of a charged particle in the exponentially varying magnetic field. The results have been presented as a function of temperature for different parameters. We have found that the entropy is increased with the temperature, and the specific heat shows a peak structure in the presence of a non-uniform magnetic field and then reduces to zero. Our findings also show that the magnetic susceptibility has a peak at a certain value of the magnetic field which depends on the temperature. It is concluded that the specific heat and magnetic susceptibility increase with external magnetic field strength and different parameter values in the region of low temperature, while the mentioned quantities are decreased in high temperature regions due to increasing the occupied levels at these regions. On the other hand, the results show that no transition from positive to negative magnetic susceptibility depending on the $\alpha$ parameter when the value of external magnetic field is kept constant. In this system, magnetic susceptibility shows an antiferromagnetic phase. It starts from low values and increases up to a critical temperature and then goes down. Also, our system does not demonstrate magnetic phase transition. The specific heat and magnetic susceptibility quantities have a maximum value and then they go to a lower value at high temperatures.

\section{Acknowledgements}
We would like to thank the kind referee for positive suggestions which have improved deeply the present paper.

\newpage

\begin{figure}[htbp]
\centering
\includegraphics[height=2.5in, width=5in, angle=0]{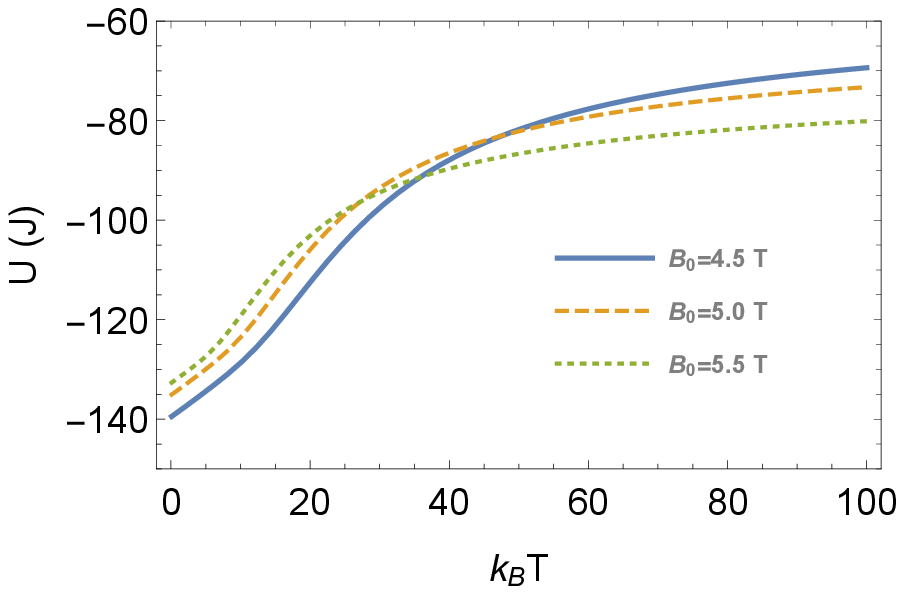}
\caption{Mean energy as a function of temperature with $\alpha=1$ and $k_y=0.1$.}
\end{figure}

\begin{figure}[htbp]
\centering
\includegraphics[height=2.5in, width=5in, angle=0]{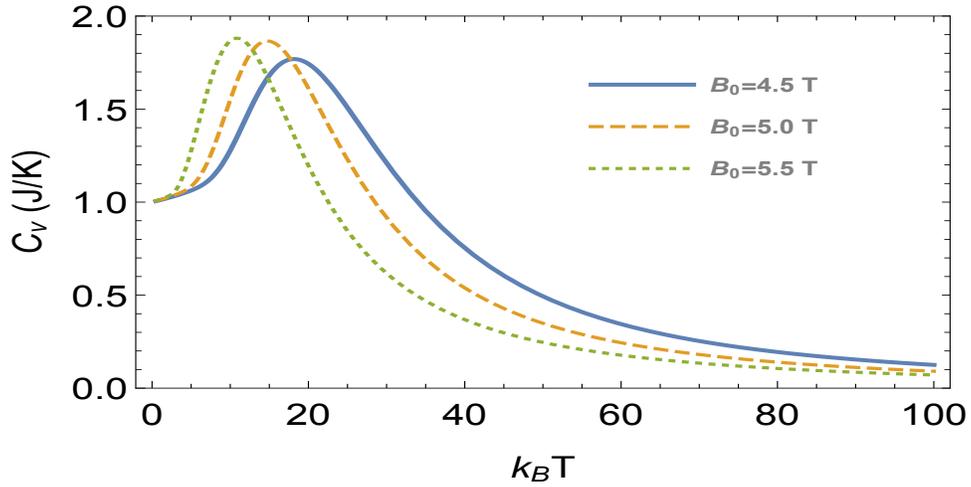}
\caption{Specific heat as a function of temperature with $\alpha=1$ and $k_y=0.1$.}
\end{figure}

\newpage

\begin{figure}[htbp]
\centering
\includegraphics[height=2.5in, width=5in, angle=0]{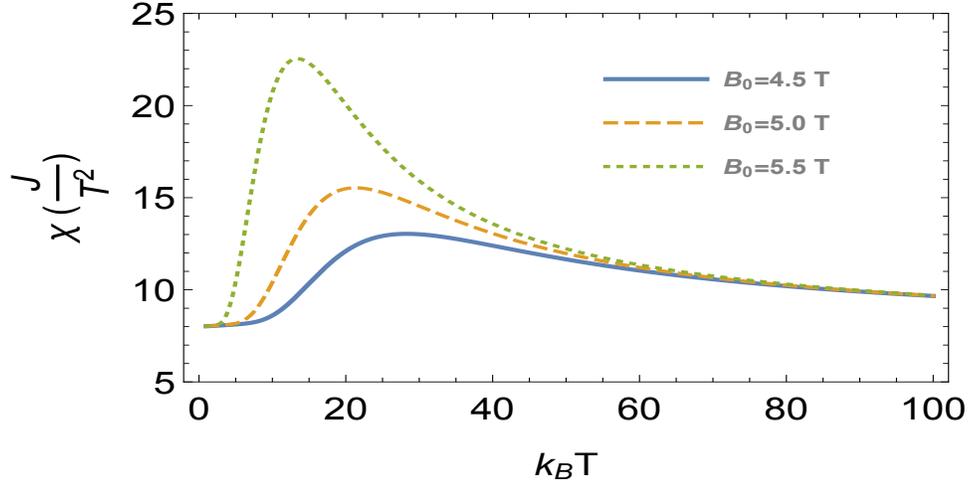}
\caption{The magnetic susceptibility as a function of temperature with $\alpha=1$ and $k_y=0.1$.}
\end{figure}

\begin{figure}[htbp]
\centering
\includegraphics[height=2.5in, width=5in, angle=0]{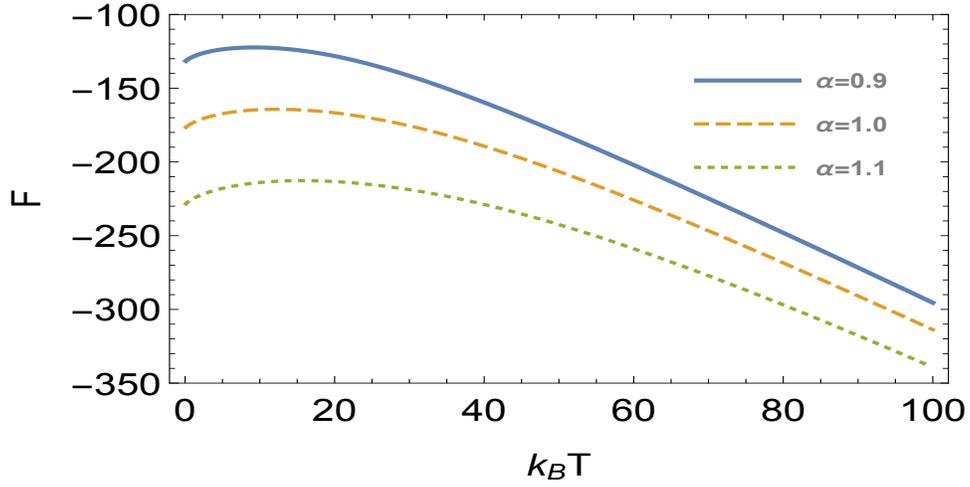}
\caption{Free energy as a function of temperature with $B=2.5\,T$ and $k_y=0.1$.}
\end{figure}

\newpage

\begin{figure}[htbp]
\centering
\includegraphics[height=2.5in, width=5in, angle=0]{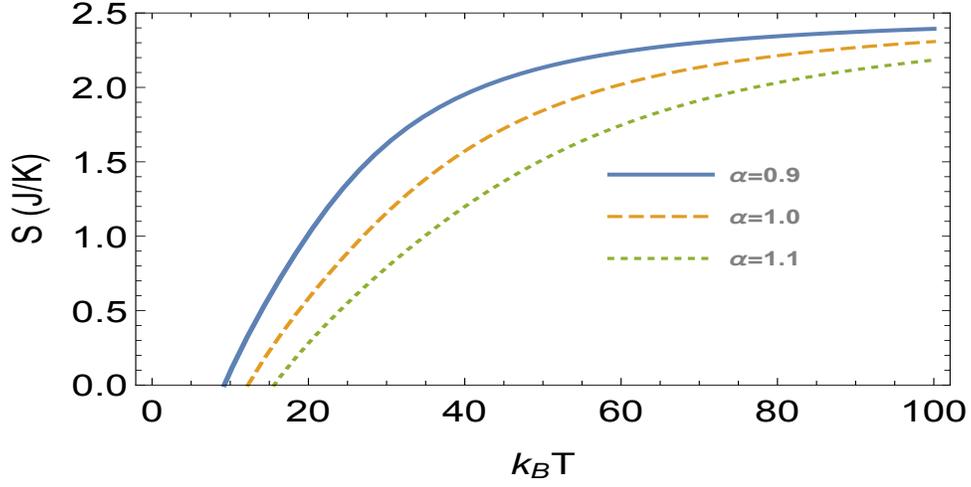}
\caption{Entropy as a function of temperature with $B=2.5\,T$ and $k_y=0.1$.}
\end{figure}

\begin{figure}[htbp]
\centering
\includegraphics[height=2.5in, width=5in, angle=0]{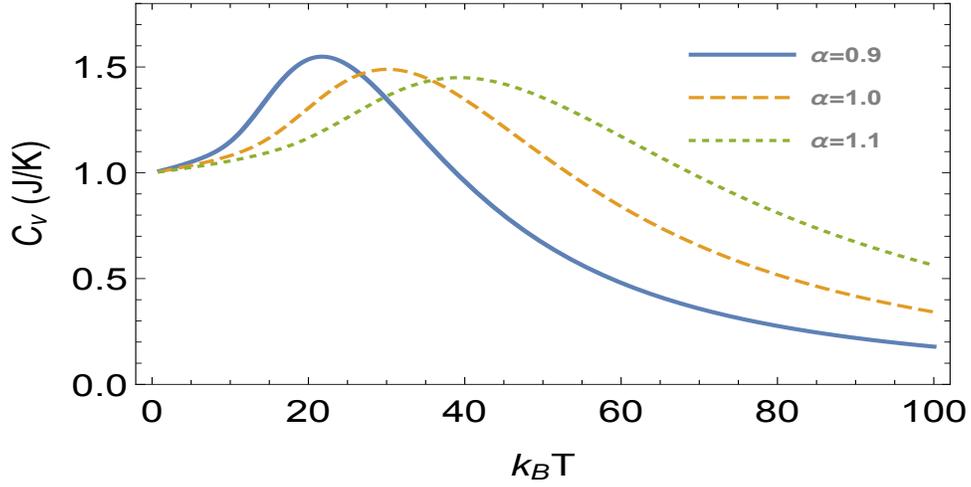}
\caption{Specific heat as a function of temperature with $B=2.5\,T$ and $k_y=0.1$.}
\end{figure}

\begin{figure}[htbp]
\centering
\includegraphics[height=2.5in, width=5in, angle=0]{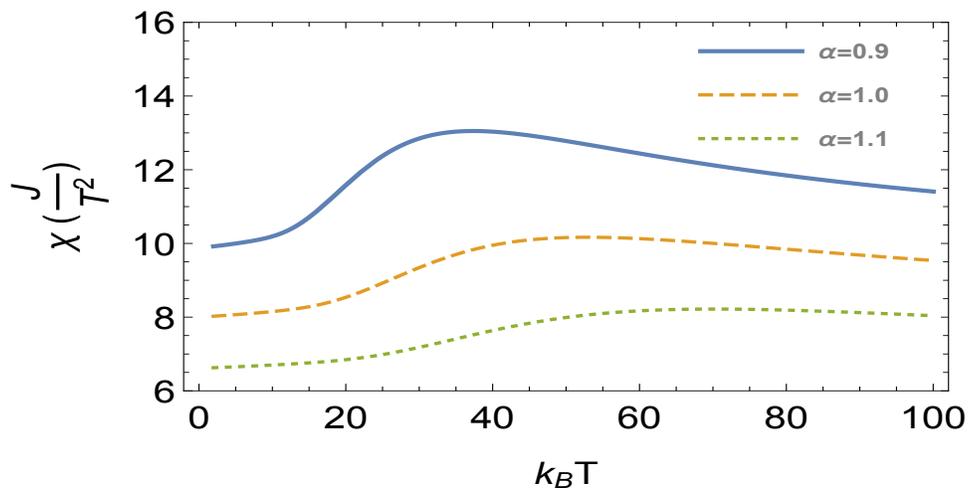}
\caption{The magnetic susceptibility as a function of temperature with $B=2.5\,T$ and $k_y=0.1$.}
\end{figure}

\end{document}